\documentclass[twocolumn,showpacs,prl]{revtex4}

\usepackage{amsmath} 
\usepackage{amssymb} 
\usepackage{graphicx}
\usepackage{psfrag}

\newcommand{\Z}{\mathbf{Z}}

\newcommand{\id}{\mathbf{1}}
\newcommand{\prima}{^\prime}
\newcommand{\primas}{^{\prime\prime}}

\newcommand{\sset}[1]{\{#1\}}

\newcommand{\pauli}{\mathcal{P}}
\newcommand{\gauge}{\mathcal{G}}
\newcommand{\stab}{\mathcal{S}}

\newcommand{\cent}{\mathcal{Z}_\gauge}

 \newcommand{\rr}{\mathrm{r}}
 \newcommand{\rg}{\mathrm{g}}
 \newcommand{\rb}{\mathrm{b}}

\begin{document}

\title{Clifford Gates by Code Deformation}

\author{H.~Bombin}
\affiliation{Perimeter Institute for Theoretical Physics, 31 Caroline St. N., Waterloo, ON, N2L 2Y5, Canada}

\begin{abstract}

Topological subsystem color codes add to the advantages of topological codes an important feature: error tracking only involves measuring 2-local operators in a two dimensional setting. Unfortunately, known methods to compute with them were highly unpractical. We give a mechanism to implement all Clifford gates by code deformation in a planar setting. In particular, we use twist braiding and express its effects in terms of certain colored Majorana operators.

\end{abstract}


\maketitle

Scalable quantum computation will not be realized unless its most fundamental problem can be addressed: decoherence. From a theoretical perspective there exist good reasons to believe that low enough levels of noise can be managed \cite{shor:1996:ftqc, aharonov:1997:ftqc}, but in practice fault-tolerance will not be experimentally achieved without proposals involving realistic assumptions. In many situations such an assumption is the locality of interactions, in a geometrical sense. Among the theoretical approaches to this problem, topological quantum codes stand out for their elegance and flexibility \cite{kitaev:2003:ftanyons}. Recently a new class of topological codes was developed \cite{bombin:2010:subsystem} that maximizes locality: only 2-local measurements are needed to keep track of errors. However, no practical way to compute with these codes was available. This paper offers one, using twists, a tool recently introduced in \cite{bombin:2010:twist}.

An \emph{error correcting code}~\cite{shor:1995:scheme,steane:1996:error} protects quantum information from decoherence by means of redundancy. This is achieved by selecting a set of commuting observables, called \emph{check operators}, that are initialized with a specific set of values, defining a \emph{code subspace}. The idea is that most errors affect the expected value of check operators, so that a repeated measurement of check operators potentially allows to keep track of errors.

A main feature of \emph{topological codes}{\cite{kitaev:2003:ftanyons} is that check operators are geometrically local. Qubits are arranged in lattices of a given dimension in such a way that the support of any check operator only involves a few neighboring qubits. The size of the lattice is arbitrary, and in the limit of large lattices topological codes show a distinctive behaviour, the appearance of an error threshold~\cite{dennis:2002:tqm} for local error models. For noise levels below this threshold, error correction is almost perfect for large lattices. In particular, the failure probability decreases exponentially with the system size~\cite{dennis:2002:tqm}. This behavior is reminiscent of a phase transition, and indeed error correction can be rephrased in terms of a statistical mechanical model  \cite{dennis:2002:tqm, katzgraber:2009:cc, bombin:2010:subsystem}. This connection has led to fast decoding algorithms~\cite{duclos:2010:fast}. The threshold persists in the presence of qubit losses\cite{stace:2010:error}.

An alternative way to get local codes is to consider \emph{subsystem codes} \cite{kribs:2005:unified}, in which only a subsystem of the code subspace is used, so that errors that do not affect this subsystem are irrelevant. The trick is that sometimes this makes it possible to split the measurement of check operators into local measurements \cite{poulin:2005:stabilizer}. An example of this are Bacon-Shor codes~\cite{bacon:2006:operator}, which take locality to its extreme: 2-local measurements are enough. Interestingly, two-dimensional stabilizer subsystem codes \cite{bravyi:2010:subsystem} do not suffer from the same constraints as conventional two-dimensional stabilizer codes \cite{bravyi:2010:tradeoffs}.

In general local subsystem codes do not enjoy the nice features of topological codes discussed above, but a recently introduced class of stabilizer subsystem codes does incorporate them \cite{bombin:2010:subsystem}. We will call them \emph{topological subsystem color codes} (TSCC) to distinguish them from other potential topological subsystem codes. As Bacon-Shor codes, TSCCs only require 2-local measurements in a 2D setting but, unlike them, they posses a threshold in the large lattice limit and allow the use of error correction techniques that are specific of topological codes.

In the realm of topological codes there are two main approaches to computation. One is the use of transversal gates \cite{dennis:2002:tqm, bombin:2006:2dcc, bombin:2007:3dcc}. The second, more specifically suited to topological codes, is \emph{code deformation} \cite{dennis:2002:tqm,raussendorf:2007:deformation,bombin:2009:deformation}. This technique plays with the possibility of changing the geometry of the code over time and allows not only to implement gates but also to initialize and measure encoded qubits in a protected way. As long as the encoded operators that require protection remain global with respect to the geometry change. code deformations can be performed simply by measuring the check operators that correspond to the new geometry \cite{bombin:2009:deformation}.

In the case of TSCCs transversal gates do not seem to be of any use, and code deformation faces an important difficulty: TSCCs do not allow \cite{bombin:2010:subsystem} the introduction of boundaries \cite{freedman:2001:planar}. Without boundaries the codes are not planar and code deformations become impractical.

This paper demonstrates how to overcome this difficulty using \emph{twists}, a tool recently introduced \cite{bombin:2010:twist} in the context of two-dimensional topologically ordered models \cite{wen:1989:to}.  These physical systems, closely related to topological codes, exhibit excitations called \emph{anyons} \cite{wilczek:1990:anyons} that interact topologically. Sometimes there exist symmetries that permute the topological charges leaving the physics unchanged \cite{beigi:2010:symmetry}. Twists are topological defects that transform anyons that encircle them according to such a symmetry. They can improve the computational power of topologically ordered systems \cite{bombin:2010:twist}.

As it will become clear, twists offer a natural way to make topological subsystem codes planar. Moreover, all Clifford gates \cite{note:clifford} can be performed by creating, moving and annihilating twists. This is enough to perform many relevant quantum information protocols. Most importantly, it gives rise to universal quantum computation through the distillation of so called magic states \cite{bravyi:2005:universal}.

\paragraph*{Stabilizer subsystem codes---}

In stabilizer codes \cite{gottesman:1996:stabilizer} the system is a collection of $n$ qubits and the code subspace is defined in terms of a subgroup $\stab$ of the Pauli group $\pauli_n:=\langle i\id, X_1, Z_1, \dots, X_n, Z_n\rangle$ \cite{note:pauli}, called stabilizer, such that $-\id\not\in\stab$. Encoded states are those with $\langle s\rangle=1$ for any $s\in \stab$, and thus the elements of any generating set of $\stab$ can be used as check operators. To encode in a subsystem only, we can choose any subgroup $\gauge\subset \pauli_n$, called gauge group, that has $\stab$ as its center, up to phases \cite{poulin:2005:stabilizer}. This splits the code into a logical subsystem, where $\gauge$ acts trivially, and a gauge subsystem, where $\cent$, the centralizer of $\gauge$ in $\pauli_n$, acts trivially. The quotient $\cent/\stab$ naturally provides Pauli operators on logical qubits, because the corresponding equivalence in $\cent$, denoted $\equiv$ below, implies an equivalent action on encoded states.

\paragraph*{Topological subsystem color codes---}

These stabilizer subsystem codes were introduced in \cite{bombin:2010:subsystem}, but here we will modify then slightly to accommodate the concept of twist. The starting point is any three-valent \cite{note:valence} lattice $\Lambda$ on an oriented closed surface. Expanding vertices into triangles and duplicating the existing links, as in Fig.~\ref{fig:lattice}(a), produces a new lattice $\bar\Lambda$. We place a qubit at each vertex of $\bar \Lambda$, and define $\gauge$ in terms of a set of 2-local generators, two per qubit. The neighborhood of any vertex can be depicted as in Fig.~\ref{fig:lattice}(a). But this can be done in two ways, and we fix an orientation so that a given side of the lattice is facing us and there is a unique way. Then to each vertex we attach the operators $Z_1Z_2$ and $Y_2X_3$ according to the labeling in the figure.

To characterize $\cent$ it is convenient to represent Pauli operators  graphically. For each vertex of $\bar\Lambda$, let us `shade' a portion of the neighboring triangles and links in one of the four ways shown in Fig.~\ref{fig:lattice}(b). We relate such a shading $\gamma$ to a Pauli operator $P_\gamma:=\bigotimes_i \sigma_i$ with $\sigma_i=\id, X,Y,Z$ as indicated in the figure. We say that $\gamma$ is closed if each link and triangle is either entirely shaded or unshaded, so that $P_\gamma\in\cent$ if and only if $\gamma$ is closed.

\begin{figure}
\psfrag{(a)}{(a)}
\psfrag{(b)}{(b)}
\psfrag{(c)}{(c)}
\psfrag{(d)}{(d)}
\psfrag{(e)}{(e)}
\psfrag{1}{\footnotesize$1$}
\psfrag{2}{\footnotesize$2$}
\psfrag{3}{\footnotesize$3$}
\psfrag{R}{\footnotesize$\rr$}
\psfrag{G}{\footnotesize$\rg$}
\psfrag{B}{\footnotesize$\rb$}
\psfrag{r}{\scriptsize$\rr$}
\psfrag{g}{\scriptsize$\rg$}
\psfrag{b}{\scriptsize$\rb$}
\psfrag{I}{\footnotesize $\id$}
\psfrag{X}{\footnotesize $X$}
\psfrag{Y}{\footnotesize $Y$}
\psfrag{Z}{\footnotesize $Z$}
\includegraphics[width=8.5cm]{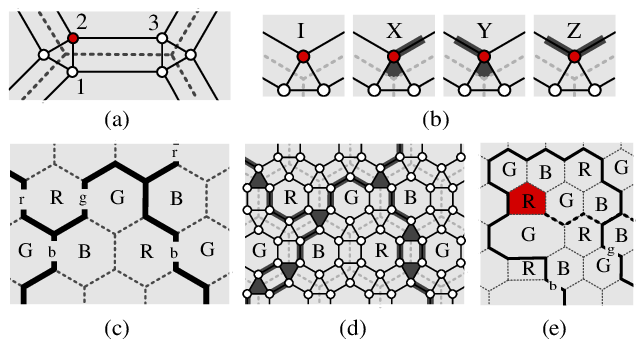}
\caption{
(a) The three-valent lattice $\Lambda$, shown dashed, gives rise to the lattice $\bar\Lambda$, in solid lines. 
(b) The correspondence between single-qubit Pauli operators and shadings.
(c) A lattice $\Lambda$ with three colorable faces, and a string-net $\gamma$.
(d) The corresponding lattice $\bar\Lambda$ and closed shading $\gamma$.
(e) A twist, a face with an odd number of links, spoils three-colorability along the dashed line. A string enclosing it changes its label.
}
\label{fig:lattice}
\end{figure}

\paragraph*{Three-colorability---}

Suppose that the faces of $\Lambda$ can be colored as red ($\rr$), green ($\rg$) and blue ($\rb$) in such a way that neighboring faces have different color, as in \cite{bombin:2010:subsystem, note:unitary}, see Fig.~\ref{fig:lattice}(c). Let us give to the set $C:=\sset{e,\rr,\rg,\rb}$ the group structure of $\Z_2\times\Z_2$, with $e$ the unit. Then closed shadings as the one in Fig.~\ref{fig:lattice}(d) are in a one-to-one correspondence with \emph{string-nets} \cite{levin:2005:stringnet} as the one in Fig.~\ref{fig:lattice}(c). String-nets are labelings of the links of $\Lambda$ with the elements of $C$ such that the product of the labels of the links meeting at any vertex is trivial. We depict strings only at those links with non-trivial labels, so that the only possible branching point involves three strings of different colors. To clarify the correspondence, let $\gamma$ be a closed shading. Each link $l$ in $\Lambda$ gives rise to two parallel links $l_1$, $l_2$ in $\bar\Lambda$, which can be labeled with the colors $c_1$, $c_2$ of their corresponding faces in $\Lambda$. The string-net is obtained by attaching to $l$ the value $l^\gamma:=l_1^\gamma l_2^\gamma$, with $l_i^\gamma=c_i$ if $l_i$ is shaded, $l_i^\gamma = e$ otherwise. In what follows we identify string-nets and closed shadings. 
As a straightforward consequence of the preceding definition, if $\gamma, \gamma\prima, \gamma\primas$ are string-nets and $L$ is the set of links in $\Lambda$, we have
\begin{equation}\label{string-nets}
P_\gamma\propto P_{\gamma\prima}P_{\gamma\primas}\quad\iff \quad\forall l\in L\quad l^\gamma=l^{\gamma\prima}l^{\gamma\primas}.
\end{equation}

We will need a generalization of string-nets in which strings are allowed to go over or under other strings, see Fig.~\ref{fig:rules}(c-e). Take any string-net $\gamma$ containing such self-crossing points. Split the shading $\gamma$ in any $p$ portions $\gamma_i$ that do not self-cross, ordered in such a way that if $i>j$ then $\gamma_i$ does not go under $\gamma_j$. Then $P_\gamma:=P_{\gamma_{r}}\cdots P_{\gamma_{1}}$, a definition independent of the $\gamma_i$ choice.

Consider now loops, string-nets without branchings. The boundary of each face $f$ of $\Lambda$ gives rise to three such loops $\gamma^c$, one per color $c$, defining the face operators $P_f^c:=P_{\gamma^c}$. 
Face operators belong to $\gauge$ and generate a valid stabilizer group $\stab$ \cite{bombin:2010:subsystem}. In particular, $-\id\not\in \stab$ because they are subject only to the global relations $\prod_f P_f^c=\id$ and the local relations $P_f^\rr P_f^\rg P_f^\rb=\id$. 
Putting together \eqref{string-nets} and the fact that the stabilizer is generated by small loops gives rises to topological equivalence rules as the ones in Fig.~\ref{fig:rules}(a-e). But \eqref{string-nets} only specify the relationship up to a sign, which has to be worked out explicitly. In particular, the deformation in (b) can be easily checked for a single plaquette, and then the other rules only have to be checked for specific examples. The rules in Fig.~\ref{fig:rules}(d,e) involving branching points do not specify the sign at all, but they will suffice. It is indeed possible to modify slightly the definition of string-nets to make all signs work nicely even when branching points are present, but this approach is not compatible with twists and in any case we will not need it. Once we have these rules for string-net manipulation, we can forget about the detailed structure of the code when analyzing encoded qubits or the effects of code deformations, as we will see.

\begin{figure}
\psfrag{(a)}{(a)}
\psfrag{(b)}{(b)}
\psfrag{(c)}{(c)}
\psfrag{(d)}{(d)}
\psfrag{(e)}{(e)}
\psfrag{(f)}{(f)}
\psfrag{(g)}{(g)}
\psfrag{=}{\large$\equiv$}
\psfrag{=1}{\large$\equiv\id$}
\psfrag{=+}{\large$\equiv \pm$}
\psfrag{=+i}{\large$\equiv \pm i$}
\psfrag{=-}{\large$\equiv -$}
\psfrag{=i}{\large$\equiv i$}
\psfrag{=1=i}{\large$\equiv \id\equiv i$}
\psfrag{c}{\scriptsize $c$}
\psfrag{p}{\scriptsize $cc\prima$}
\psfrag{s}{\scriptsize $\,c$}
\psfrag{d}{\scriptsize $c\prima$}
\includegraphics[width=8.5cm]{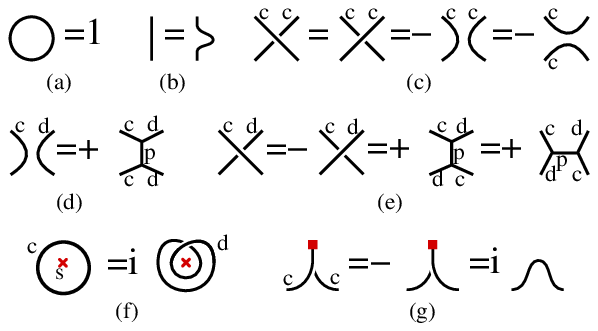}
\caption{
Equivalence rules for string-net operators $P_\gamma$ in terms of the corresponding string-nets $\gamma$. Crosses mark twists, squares mark common endpoints, $\pm$ signs are unrelated, $c,c\prima \in \sset{\rr, \rg,\rb}$ label the strings and $c\neq c\prima$.
}
\label{fig:rules}
\end{figure}

\paragraph*{Twists---}

Consider again the general case of non-three-colorable lattices, but restricted to lattices in the sphere. These can in practice be realized in the plane, by introducing arbitrary boundaries that only affect error correction.  The three-colorability of $\Lambda$ is spoiled by faces with an odd number of links, see Fig.~\ref{fig:lattice}(e), which we call \emph{twists}, as opposed to \emph{regular} faces. Since all the properties of the code will turn out to have a topological character, we will suppose without loss of generality that there are $t$ twists placed along a line $T$, see Fig.~\ref{fig:twists}(b), and that the coloring is such that faces of the same color only meet at this line.

String-nets are still useful if labels are carefully managed, as now they can change along a string. If a string encloses a twist, 
moving counterclockwise along it induces a label permutation. Twists are thus labeled by the group of symmetries of $C$, $\Gamma\simeq S_3$ . 
In particular $\Gamma=\sset{e, \zeta_+, \zeta_-, \sigma_\rr, \sigma_\rg, \sigma_\rb}$, with $\zeta_+$ the cycle $(\rr\,\,\rg\,\,\rb)$, $\zeta_-$ its inverse and $\sigma_c$ the transposition with $\sigma_c(c)=c$. A face $f$ with an odd number of links gives rise to a transposition $\sigma_c$, and the face operator $P_f^c$ is well defined and belongs to the center of $\gauge$ \cite{note:3-local}. 

Ordering twists from left to right, let $\pi_i\in \Gamma$ be the effect of the $i$-th twist. Although twists only produce transpositions, putting them together produces composite twists with arbitrary labels. Indeed, the combined effect of the $i$-th and $(i+1)$-th twists is $\pi_i\circ\pi_{i+1}$. Because we work in a sphere, the net effect of twists must be trivial, $\pi_1\circ\pi_2\circ\cdots \circ\pi_{t}=e$. In particular, $t$ is even because transpositions are odd permutations. It is convenient to label twists with colors $c_i$ setting  $\sigma_{c_i}:=\pi_i$.

\paragraph*{Stabilizer and logical Pauli operators---} 

Let $\stab$ be the group generated by all face operators. In the presence of twists there are no global relations between face operators unless all twists have the same color $c$. In that case the relation $\prod_f P_f^c=\id$ survives. It follows that $-\id\not \in\stab$ and we choose any stabilizer $\stab\prima$ with $\stab\subset \stab\prima$. We will prove below that $\stab\prima=\stab$, but for the moment we can already state the equivalences up to $\stab$ of Fig.~\ref{fig:rules}(f). 

Taking away from the lattice the twist faces and the links lying on $T$ produces a simpliy connected surface, a disc. 
Then on it we can apply the rules (a-e), so that any element of $\cent$ can be related to a string-net as the one in Fig.~\ref{fig:twists}(a), with strings living only on $T$ and the boundaries of twists. 
Up to the twist face operators, these string-nets are characterized by the labels $d_i\in C$ of the strings connecting twists, and thus we label them as $\gamma_T^{d_1,\dots,d_{t-1}}$. 
For this labeling to be well defined, we assume that it corresponds to the coloring of a given side of $T$. Then branching rules impose the constraints $d_{i-1}d_i\in\sset{e,c_i}$,  for $i=1,\dots,t$ with $d_0:= d_t:=e$.

Our next goal is to understand $\cent$. Consider firstly the case where all twists have the same color. 
Set for concreteness $c_i=\rr$ and consider loops enclosing an even number of twists. In particular, Let $\gamma_{i,j}^c$ denote a loop enclosing the twists $i$ to $j$, $1 \leq i<j\leq t$ and $j-i$ even, with $c$ its color at some arbitrary but fixed point along the loop. If $i$ is even the color is the same in the whole loop, but if it is odd and $c=\rg$ or $c=\rb$ it flips when crossing $T$.
Define the operators 
\begin{equation}\label{logical}
\hat X_i:=P_{\gamma^b_{2i,2i+1}}, \quad \hat Z_i:=P_{\gamma^b_{2i+1,t}}, \qquad  i=1,\dots, k
\end{equation}
with $k=t/2-1$. These operators are the logical $X$ and $Z$ Pauli operators for $k$ logical qubits, as they satisfy the corresponding commutation rules. Notice that the string-nets $\gamma_T^{d_1,\dots,d_{t-1}}$ have $d_i=e,\rr$, and that $Q:=P_{\gamma_T^{d_1,\dots,d_{t-1}}}$ commutes with all the operators \eqref{logical} only if $d_i=e$ for all $i$, giving $Q\equiv \id$, or if $d_i=\rr$ for all $i$, giving $Q \equiv P_{\gamma^\rb_{1,t}}\equiv \id$. Therefore, $\cent/\stab\simeq \pauli_k$, showing that there are exactly $k$ encoded qubits and $\stab=\stab\prima$. Any nontrivial element of $\cent$ must connect at least two twists and thus local errors cannot affect encoded qubits, a consequence of their topological nature \cite{bombin:2010:subsystem}.

When not all twists have the same label there is a trick to simplify the analysis. Changing the order in which the line $T$ visits twists changes their labeling, and there is always a choice such that $c_i=\rr$ for $1\leq i \leq t\prima$ and $c_i=\rg$ for $t\prima<i\leq t$ for some even integer $t\prima$ ---this might not be obvious yet, but it will be below after the closely related twist braiding is analyzed---. Since $d_t=e$ and $P_{\gamma^c_{1,t\prima}}\equiv P_{\gamma^c_{t\prima+1,t}} \equiv \id$, each block can be treated separately exactly as above giving $k=t/2-2$.

\paragraph*{Colored Majorana operators---} 

Twists can be moved around with the code deformation technique described in \cite{bombin:2009:deformation}. This will produce topologically protected gates on encoded qubits that can be understood most simply by following the evolution of logical Pauli operators, which is easier for loops. As a twist moves, a loop $\gamma$ will be dragged to a new loop $\gamma\prima$, and $P_\gamma$ maps to $P_{\gamma\prima}$.

To give a convenient description of twist braiding, we will use \cite{bombin:2010:twist} a set of open string operators. They do not belong to $\cent$, but any product of them that forms a (generalized) loop does belong to $\cent$. It is important that all the strings have the same endpoint geometry, so that the simple rules in Fig.~\ref{fig:rules}(g) can be applied.

\begin{figure}
\psfrag{(a)}{(a)}
\psfrag{(b)}{(b)}
\psfrag{(c)}{(c)}
\psfrag{(d)}{(d)}
\psfrag{(e)}{(e)}
\psfrag{s1}{\scriptsize $c_1$}
\psfrag{s2}{\scriptsize $c_2$}
\psfrag{s3}{\scriptsize $c_3$}
\psfrag{s4}{\scriptsize $c_4$}
\psfrag{s5}{\scriptsize $c_5$}
\psfrag{sr}{\scriptsize $\,\rr$}
\psfrag{sg}{\scriptsize $\,\rg$}
\psfrag{sb}{\scriptsize $\,\rb$}
\psfrag{rr}{\scriptsize $\rr$}
\psfrag{rg}{\scriptsize $\rg$}
\psfrag{rb}{\scriptsize $\rb$}
\psfrag{cr}{\scriptsize $\gamma_{1,3}^\rr$}
\psfrag{g1}{\scriptsize $\gamma_5$}
\psfrag{g2}{\scriptsize $\gamma_7$}
\psfrag{g3}{\scriptsize $\gamma_9$}
\psfrag{d1}{\scriptsize $d_1$}
\psfrag{d2}{\scriptsize $d_2$}
\psfrag{d3}{\scriptsize $d_3$}
\psfrag{d4}{\scriptsize $d_4$}
\psfrag{c}{\scriptsize $c$}
\psfrag{cp}{\scriptsize $c\prima$}
\psfrag{s}{\scriptsize $\sigma_c(c\prima)$}
\includegraphics[width=8.5cm]{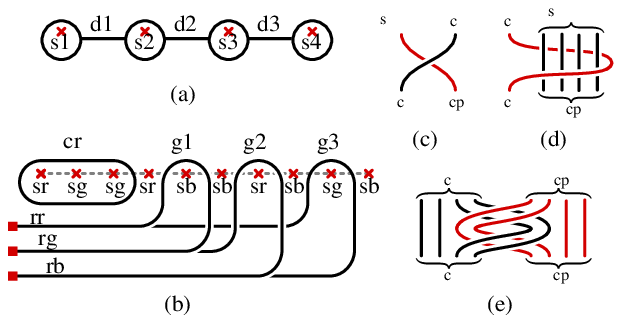}

\psfrag{s}{\footnotesize$\sigma_c(c\prima)$}
\psfrag{c}{\footnotesize$c$}
\psfrag{cp}{\footnotesize$c\prima$}
\psfrag{j}{\footnotesize $j$}
\psfrag{jp}{\footnotesize $j+1$}

\caption{
(a) A string-net $\gamma_T^{d_1,d_2, d_3}$ in a geometry with four twists.
(b) Ten twists lying on the $T$ line, dashed. A loop enclosing three twists and three open string operators $\gamma_i$. These have endpoints in three fixed points, one per color. Notice how $\gamma_i$ connects the endpoints with color $\zeta_+(c_i)$ and $\zeta_-(c_i)$ without self-intersecting.
(c-e) Twist braiding processes. Time flows upwards. 
(c) An elementary braiding of two twists. 
(d) Changing the color of an encoded qubit using another twist. 
(e) An entangling gate between two encoded qubits if $c\neq c\prima$.
}
\label{fig:braiding
}
\label{fig:twists}
\end{figure}

So let us attach an open string $\gamma_i$ to the $i$-th twist as in Fig.~\ref{fig:twists}(b), and set $k_i:=P_{\gamma_i}$. The self-adjoint operators $k_i$ satisfy $k_i^2=1$ and, for $i<j$,
\begin{equation}\label{commutation}
   k_i k_j = \left\{
     \begin{array}{ll}
       k_j k_i &\,\text{ if } c_i=\zeta_+(c_j),\\
       -k_j k_i & \,\text{ otherwise.}
     \end{array}
   \right.
\end{equation}
Since the $k_i$ operators on twists of the same color behave like Majorana operators, we call them colored Majorana operators. The logical operators \eqref{logical} can be recovered as 
\begin{equation}\label{logical2}
\hat X_j \equiv -i k_{2j}k_{2j+1},\quad \hat Z_j \equiv (-i)^{\frac t 2-j} k_{2j+1}k_{2j+2}\cdots k_t.
\end{equation}

Braiding the $j$-th and $(j+1)$-th twist as in Fig.~\ref{fig:twists}(c) changes the color of the twists to $c_{j}\prima = \sigma_{c_j}(c_{j+1})$ and $c_{j+1}\prima = c_j$.
As the twists move, string operators are dragged \cite{bombin:2010:twist}. Using the transformation rules, one can relate this dragged string operators to the standard ones. The net effect of the braiding turns out to be
\begin{equation}\label{braiding}
   k_j \rightarrow k_{j+1}, \,\,\,
   k_{j+1} \rightarrow \left\{
     \begin{array}{ll}
       - k_j &\text{ if } c_j=c_{j+1},\\
        i k_j k_{j+1} &\text{ if } c_j=\zeta_-(c_{j+1}),\\
       -k_j k_{j+1} & \text{ otherwise.}
     \end{array}
   \right.
\end{equation}
Remarkably, the rules for braiding twists of the same color reduce to the ones appearing in \cite{bombin:2010:twist}, which in turn allows us to use the results in \cite{bravyi:2006:ising}. From \eqref{logical2} and \eqref {braiding} it follows that the $k_j$-s suffice to describe $\cent/\stab$.

\paragraph*{Clifford gates---} 

The transformation rules \eqref{braiding} are rich enough to reproduce the whole Clifford group \cite{note:clifford}. Indeed, it follows from \cite{bravyi:2006:ising} that all single qubit Clifford gates, initializations and measurements included, can be realized by encoding each qubit in four subsequent twists of the same color, working in the subspace $\langle k_j k_{j+1} k_{j+2} k_{j+3}\rangle=-1$ and choosing
\begin{equation}\label{logical3}
\hat X \equiv -i k_{j}k_{j+1},\qquad \hat Z \equiv -i k_{j+1}k_{j+2}. 
\end{equation}
To show that all Clifford gates can be implemented it suffices to exhibit the realization of an entangling gate. The braiding of Fig.~\ref{fig:twists}(e) on a pair of encoded qubits of \emph{different} colors produces such an entangling gate:
\begin{alignat}{2}
&\hat X_1 \rightarrow \hat X_1, \qquad  &\hat Z_1 \rightarrow \hat X_2 \hat Z_1, \nonumber\\
&\hat X_2 \rightarrow \hat X_2, \qquad  &\hat Z_2 \rightarrow \hat X_1 \hat Z_2.
\end{alignat}
Since the color of the four twists composing a qubit can be changed anytime by braiding a suitable twist around them, as in Fig.~\ref{fig:twists}(d), we have the desired result.

\paragraph*{Conclusions and outlook---}

Topological subsystem color codes are optimal in terms of their locality properties, in the sense that they only require 2-local measurements for error tracking. We have shown that they also achieve the maximum computational capabilities within topological code deformation, since it is not possible to go beyond Clifford gates (as logical Pauli operators are always mapped to other logical Pauli operators). This makes them a very interesting alternative within the realm of topological codes, which have many advantages in terms of locality, error thresholds and error correction procedures.

The mechanism that we have used are twists, which are based on anyon symmetries \cite{bombin:2010:twist}. In the case of topological subsystem color codes, the relevant anyon model, as dictated by the properties of string operators, contains three fermionic charges, one per color, with semionic mutual statistics \cite{bombin:2010:subsystem}. The symmetry of this anyon model is given by the symmetric group $S_3$. A remarkable aspect of the corresponding generalized topological charges \cite{bombin:2010:twist}, which include the twist labels, is that their fusion rules are non-commutative, a consequence of $S_3$ being non-abelian. This clearly shows that twists cannot be regarded as anyons, at least not directly.

The essence of the technique used here can be adapted both to color codes \cite{bombin:2006:2dcc} and to Majorana color codes \cite{bravyi:2010:majorana}. The reason for this is that the corresponding anyon models contain three charges with semionic mutual statistics and symmetry $S_3$. In the case of regular color codes there are many more possible symmetries to play with. Among them, the $S_3$ symmetry related to the exchange of the $X$, $Y$ and $Z$ type check operators is particularly convenient, as it does not involve playing with geometry. This symmetry, however, does not correspond to the one studied here and so would require a separate analysis of its computational power.

\begin{acknowledgments} 


This work was supported with research grants QUITEMAD S2009-ESP-
1594, FIS2009-10061 and UCM-BS/910758.

\end{acknowledgments}

\bibliography{refs,comments}

\begin{thebibliography}{34}
\bibitem[{not({\natexlab{a}})}]{note:clifford}
\bibinfo{note}{The Clifford group has as elements those unitary operators
  mapping the Pauli group to itself under conjugation.}
\bibitem[{not({\natexlab{c}})}]{note:valence}
\bibinfo{note}{A lattice is $n$-valent if $n$ links meet at each vertex.}
\bibitem[{not({\natexlab{d}})}]{note:unitary}
\bibinfo{note}{The gauge generators in \cite{bombin:2010:subsystem} are not the
  same used here, but they are equivalent up to qubit rotations. The reason for
  the change is that the gauge operators defined in
  \cite{bombin:2010:subsystem} are not suitable for plaquettes with an odd
  number of edges.}
\bibitem[{not({\natexlab{e}})}]{note:3-local}
\bibinfo{note}{Unlike regular face operators, such twist face operator cannot
  be recovered from measurements of the 2-local gauge operators. Fortunately,
  it is not necessary to be able to measure all face operators in order to keep
  track of errors. Thus, as long as twists are far away from each other,
  2-local measurements are still enough.}
\end{thebibliography}

\end{document}